\documentclass [3p, 12pt, sort&compress]{elsarticle}
\usepackage[numbers]{natbib}
\usepackage{graphicx}
\usepackage[small]{subfigure,epsfig}
\usepackage{indentfirst}
\usepackage{amsmath,latexsym,enumerate}
\usepackage{amssymb}
\usepackage{amsfonts}

\usepackage{amsthm}

\theoremstyle{plain} \newtheorem{theorem}{Theorem}

\numberwithin{equation}{section} \numberwithin{lemma}{section} \numberwithin{theorem}{section}
\begin{document}
\begin{frontmatter}

\title{Invariant algebraic curves for Li\'{e}nard dynamical systems revisited}

\author{Maria V. Demina}

\address{Department of Applied Mathematics, National Research Nuclear University "MEPhI", 31 Kashirskoe Shosse,
115409 Moscow, Russian Federation, mvdemina@mephi.ru}

\begin{abstract}

A novel algebraic method for finding invariant algebraic curves for a polynomial vector field in $\mathbb{C}^2$ is introduced.
The structure of  irreducible invariant algebraic curves for   Li\'{e}nard dynamical systems $x_t=y$, $y_t=-f(x)y-g(x)$ with $\deg g=\deg f+1$ is obtained.
It is shown that there exist Li\'{e}nard systems that possess more complicated invariant algebraic curves than it was supposed before.
 As an example, all irreducible invariant algebraic curves for the Li\'{e}nard differential system with $\deg f=2$ and $\deg g=3$ are obtained. 
 All these results seem to be new.

\end{abstract}


\end{frontmatter}

\section{Introduction}\label{Introduction}

The integrability and limit cycles of famous Li\'{e}nard dynamical systems
\begin{equation}
\begin{gathered}
 \label{Lienard_gen}
 x_t=y,\quad  y_t=-f(x)y-g(x)
\end{gathered}
\end{equation}
have been intensively studied in recent years \cite{Zoladec01, Llibre05, Llibre04, Llibre06, Llibre07, Ignatyev01, Amelkin01, Cherkas01, Cheze01, Liu01}.
In this article we suppose that $f(x)$ and $g(x)$ are
polynomials
\begin{equation}
\begin{gathered}
 \label{Lienard_fg}
f(x)=f_0x^m+\ldots+f_m, \quad g(x)=g_0x^{m+1}+\ldots+g_{m+1},\quad f_0\neq0,\quad g_0\neq0
\end{gathered}
\end{equation}
with coefficients in the field $\mathbb{C}$.

An algebraic curve $F(x,y)=0$, $F(x,y)\in \mathbb{C}[x,y]\setminus\mathbb{C}$
is  an invariant algebraic curve  (or a Darboux polynomial) of the polynomial dynamical  system
\begin{equation}
\begin{gathered}
 \label{DS}
 x_t=P(x,y),\quad y_t=Q(x,y);\quad P(x,y),\quad Q(x,y)\in \mathbb{C}[x,y]
\end{gathered}
\end{equation}
if it satisfies the following equation
\begin{equation}
\begin{gathered}
 \label{Inx_Eq}
P(x,y)F_x+Q(x,y)F_y=\lambda(x,y) F,
\end{gathered}
\end{equation}
where $\lambda(x,y)\in \mathbb{C}[x,y]$ is  a polynomial called the cofactor of the invariant algebraic curve $F(x,y)$.
By $\mathbb{C}[x,y]$ we denote the ring of polynomials in the variables $x$ and $y$
with coefficients in the field $\mathbb{C}$.

It can be observed that an invariant
algebraic curve of dynamical system \eqref{DS} is formed by solutions of the latter. A solution of system \eqref{DS}
has either empty intersection with the zero set of $F$ or it is entirely contained in $F= 0$.
Existence of invariant algebraic curves is a substantial measure of integrability, for more details see, for instance \cite{Llibre01, Goriely, Zhang, Gine02}.
It is an important problem to classify all irreducible invariant algebraic curves of
a dynamical system.

It was stated by Hayashi \cite{Hayashi01} that if  a Li\'{e}nard dynamical system  \eqref{Lienard_gen} with $\deg g=\deg f+1$ had an invariant
algebraic curve then the latter should be of the form
$F(x,y)=y-R(x)$, where $R(x)\in\mathbb{C}[x]$. In this article we show that there exist  Li\'{e}nard  dynamical systems \eqref{Lienard_gen} with
$\deg g=\deg f+1$ that possess more complicated invariant algebraic curves. Our main results are formulated in the following theorems.

\begin{theorem}\label{T1}
Let $F(x,y)\in \mathbb{C}[x,y]\setminus\mathbb{C}$ be an irreducible  invariant algebraic curve of  Li\'{e}nard dynamical system \eqref{Lienard_gen} with $\deg g=\deg f+1$.
Then  $F(x,y)$ takes the form
\begin{equation}
\begin{gathered}
 \label{Lienard1_F}
F(x,y)=\left\{\prod_{j=1}^{N-k}\left\{y-y_j(x)\right\}\left\{y-y_N(x)\right\}^{k}\right\}_{+},
\end{gathered}
\end{equation}
where $k=0$ or $k=1$, $N\in\mathbb{N}$, and $y_1(x)$, $\ldots$, $y_{N}(x)$ are the series
\begin{equation}
\begin{gathered}
 \label{Lienard1_F_series}
(I):\,y_j(x)=q(x)+\sum_{l=0}^{\infty}c_l^{(j)}x^{-l},\quad j=1,\ldots, N-k;\\
(II):\,y_N(x)=-\frac{g_0}{f_0}x+\sum_{l=0}^{\infty}c_l^{(N)}x^{-l}.\hfill
\end{gathered}
\end{equation}
The symbol $\{W(x,y)\}_{+}$ means that we take the polynomial part of $W(x,y)$. 
The coefficients of the series of type (II) and of the polynomial 
\begin{equation}
\begin{gathered}
 \label{Lienard1_F_series_q}
q(x)=-\frac{f_0}{m+1}x^{m+1}+\sum_{l=1}^{m} q_l x^l\in\mathbb{C}[x]
\end{gathered}
\end{equation}
 are uniquely determined.
The coefficients $c_0^{(j)}$, $j=1,\ldots, N-k$ are pairwise distinct. All other coefficients
 $c_n^{(j)}$, $n\in\mathbb{N}$ are expressible via $c_0^{(j)}$, where $j=1,\ldots, N-k$.
 The corresponding product in  \eqref{Lienard1_F} is unit whenever $k=1$ and $N=1$.
\end{theorem}

Introducing the invertible  change of  variables $x=s$, $y=z-h(s)$ $\leftrightarrow$ $s=x$, $z=y+h(x)$,
where the polynomial $h(x) $ has degree $m+1$ and is defined via the relation
\begin{equation}
\begin{gathered}
 \label{Lienard1_h}
h_x=f+\varrho,\quad \varrho=-\frac{(m+1)g_0}{f_0},
\end{gathered}
\end{equation}
we obtain the following dynamical system
\begin{equation}
\begin{gathered}
 \label{Lienard1_alt_DS}
s_t=z-h(s),\quad z_t=\varrho\{z-h(s)\}-g(s).
\end{gathered}
\end{equation}
The constant $h(0)$ may be taken arbitrary. For example, we may set $h(0)=0$. There exists the one--to--one correspondence between irreducible invariant algebraic curves
$F(x,y)$ of Li\'{e}nard dynamical system  \eqref{Lienard_gen} and irreducible invariant algebraic curves
$G(s,z)$ of system~\eqref{Lienard1_alt_DS}.

\begin{theorem}\label{T2_L_alt}
Let $G(s,z)\in \mathbb{C}[s,z]\setminus\mathbb{C}$ be an irreducible invariant algebraic curve of   dynamical system \eqref{Lienard1_alt_DS}, where $h(s)$
is given by \eqref{Lienard1_h}. Then  the degree of $G(s,z)$ with respect to $s$ is either $0$, or $m+1$.
\end{theorem}
The latter theorem is very important for applications, because the bound on the degrees of irreducible invariant algebraic curves established in the theorem can be used
to find all irreducible invariant algebraic curves  of systems \eqref{Lienard_gen}, \eqref{Lienard1_alt_DS} in explicit form.

As an example, we consider the following Li\'{e}nard dynamical system
\begin{equation}
\begin{gathered}
 \label{Lienard1_DS23}
x_t=y,\quad y_t=-(\zeta x^2+\beta x+\alpha)y-(\varepsilon x^3+ex^2+\sigma x+\delta);\quad  \zeta\neq0,\quad \varepsilon\neq0.
\end{gathered}
\end{equation}
Dynamical properties of this system were studied by Cherkas and Sidorenko \cite{Cherkas01}.
The change of variables $x\mapsto A(x+B)$, $y\mapsto Ay$, $A\neq 0$ relates system \eqref{Lienard1_DS23} with its simplified version at $\zeta=3$, $e=0$.
Thus without loss of generality, we set $\zeta=3$, $e=0$.
%Other parameters are supposed to be from the field $\mathbb{C}$.

\begin{theorem}\label{T3_L23}
There exist only nine irreducible invariant algebraic curves of  Li\'{e}nard dynamical system \eqref{Lienard1_DS23} with $\zeta=3$, $e=0$, and $\varepsilon\neq0$.
They are given in table \ref{T:FH_inv}.
\end{theorem}
It seems that the classification of
irreducible invariant algebraic curves of  system \eqref{Lienard1_DS23} is presented here for the first time.
%In the next section we shall prove theorems \ref{T1}, \ref{T2_L_alt}, \ref{T3_L23} and  present
%a method, which can be used to construct invariant algebraic curves of a polynomial dynamical system.

\section{Proof of main results} \label{P}

Let us consider polynomial dynamical system \eqref{DS}. Regarding the variable $y$ as dependent and the variable $x$ as independent, we see that the function $y(x)$
satisfies the following first--order ordinary differential equation
\begin{equation}
\begin{gathered}
 \label{ODE_y}
P(x,y)y_x-Q(x,y)=0.
\end{gathered}
\end{equation}
 In what follows we suppose that  the polynomials $P(x,y)$ and $Q(x,y)$ do not have non--constant common factors.  Note that the roles of $y$ and $x$ can be changed.

A Puiseux series in a neighborhood of the point $x=\infty$ is defined as
\begin{equation}
\begin{gathered}
 \label{Puiseux_inf}
y(x)=\sum_{l=l_0}^{+\infty}b_lx^{-\frac{l}{n}}
\end{gathered}
\end{equation}
where $l_0\in\mathbb{Z}$, $n\in\mathbb{N}$. It follows from the classical results that a Puiseux series of the form \eqref{Puiseux_inf} that satisfy the equation $F(x,y)=0$,
$F(x,y)\in\mathbb{C}[x,y]$ is convergent in a neighborhood of the point $x=\infty$ (the point $x=\infty$ is excluded from domain of convergence if $l_0<0$) \cite{Walker}.
 The set of all Puiseux series of the form \eqref{Puiseux_inf} forms a field, which we denote by $\mathbb{C}_{\infty}\{x\}$. It can be easily proved that
 if $y(x)$ is  a Puiseux series satisfying the equation $F(x,y)=0$, $F_y\not\equiv0$ with $F(x,y)$ being an invariant algebraic curve    of dynamical system \eqref{DS},
 then the series $y(x)$ satisfies equation  \eqref{ODE_y}, see \cite{Demina05, Demina07}.
 All the Puiseux series that solve equation  \eqref{ODE_y} can be obtained with the help
 of the Painlev\'{e}
 methods, see, for example \cite{Goriely, Bruno02, Bruno01, Conte03, Demina01, Demina02, Demina07}.
Some methods and algorithms related to first integrals, algebraic functions, and Puiseux series are described in
\cite{Ferragut01, Giacomini01, Giacomini02, Gine01}.

\textit{Proof of theorem \ref{T1}.}  Invariant algebraic curves of of dynamical system \eqref{Lienard_gen} 
satisfy the following equation
\begin{equation}
\begin{gathered}
 \label{Lienard_gen_main_F}
yF_x-\{f(x)y+g(x)\}F_y=\lambda(x,y) F.
\end{gathered}
\end{equation}
Substituting $F=F(x)$ into this equation, we verify that there are no invariant algebraic curves that do not depend on $y$.

Let $F(x,y)$, $F_y\not\equiv0$  be an irreducible invariant algebraic curve    of dynamical system \eqref{Lienard_gen}.
The field $\mathbb{C}_{\infty}\{x\}$ is algebraically closed \cite[chapter $IV$, section 3, theorem 3.1]{Walker}.
There exists uniquely determined system of elements $y_j(x)\in \mathbb{C}_{\infty}\{x\}$
such that the following representation is valid~\cite[chapter $IV$, section 3, theorem 3.2]{Walker}:
\begin{equation}
\begin{gathered}
 \label{F_rep_0}
F(x,y)=\mu(x)\prod_{j=1}^N\{y-y_j(x)\},
\end{gathered}
\end{equation}
where $N$ is the degree of $F(x,y)$ with respect to $y$ and $\mu(x)\in \mathbb{C}[x]$. Moreover, if a non--constant polynomial $p(x)\in \mathbb{C}[x]$ does
not divide $F(x,y)$, then $F(x,y)$ has multiple factors in $\mathbb{C}[x,y]$ if and only if the equation $F(x,y)$ has multiple roots
in $\mathbb{C}_{\infty}\{x\}$ \cite[chapter $IV$, section 3, theorem 3.5]{Walker}.
Further,  the set of elements $y_n(x)\in \mathbb{C}_{\infty}\{x\}$ appearing in representation \eqref{F_rep_0} is a
subset of those satisfying equation~\eqref{ODE_y}~\cite{Demina05, Demina07}.

Balancing the higher--order terms with respect to $y$ in  equation  \eqref{Lienard_gen_main_F}, we find  $\mu(x)=\mu_0\in\mathbb{C}$.
Without loss of generality, we set $\mu_0=1$. Equation \eqref{ODE_y}  now takes the form
\begin{equation}
\begin{gathered}
 \label{Lienard_y_x}
yy_x+f(x)y+g(x)=0.
\end{gathered}
\end{equation}

There exist only two dominant balances that produce Puiseux series in a neighborhood of the point $x=\infty$. These balances and their solutions are the following
\begin{equation}
\begin{gathered}
 \label{Lienard_balances}
(I):\quad y(y_x+f_0x^m)=0,\quad y(x)=-\frac{f_0}{m+1}x^{m+1};\\
(II):\quad x^m(f_0y+g_0x)=0,\quad y(x)=-\frac{g_0}{f_0}x.\hfill
\end{gathered}
\end{equation}
In the case $(I)$ the corresponding Puiseux series has one arbitrary coefficient at $x^0$ if the compatibility
condition related to the unique Fuchs  index $l=0$ is satisfied. Let us recall that the definition of Fuchs indices depends
on the numeration of the series under consideration. In this article we use the numeration as given in \eqref{Puiseux_inf}. 
If the parameters of system \eqref{Lienard_gen}
are chosen in such a way that this condition is not satisfied, then the series of type $(I)$ does not exist.
In the case $(II)$ all the coefficients of the corresponding series are uniquely determined. Note that the Puiseux series are in fact Laurent series, they are given in
\eqref{Lienard1_F_series}.
Taking the polynomial part of representation~\eqref{F_rep_0} we obtain~\eqref{Lienard1_F}.  Since the polynomial $F(x,y)$ in ~\eqref{Lienard1_F} is irreducible, we
see that the series $y_1(x)$, $\ldots$, $y_{N-k}(x)$ should be pairwise distinct \cite{Walker}. Thus the irreducibility of $F(x,y)$  requires the coefficients
  $c_0^{(j)}$, $j=1,\ldots, N-k$
to be pairwise distinct and $k=0$ or $k=1$.
 This completes the prove.

\textit{Remark.} If the parameters of system \eqref{Lienard_gen}
are chosen in such a way that the compatibility condition for the series of type $(I)$ is not satisfied, then
the unique irreducible invariant algebraic curve takes the form $F(x,y)=y+(g_0x)/f_0-c_0^{(N)}$ provided that the series of type $(II)$ terminates.

\textit{Proof of theorem \ref{T2_L_alt}.} Let $G(s,z)$  be an irreducible invariant algebraic curve    of dynamical system \eqref{DS}.
In what follows we regard $s$ as a dependent variable and $z$ as an  independent. Our aim is to find the representation of $G(s,z)$ in $\mathbb{C}_{\infty}\{z\}$.
The equation for the function $s(z)$ reads as
\begin{equation}
\begin{gathered}
 \label{Lienard_alt_sz}
[\varrho\{z-h(s)\} -g(s)]s_z+h(s)-z=0.
\end{gathered}
\end{equation}
If the coefficient $\varrho$ is defined as in  \eqref{Lienard1_h}, then there exists only one dominant balance producing Puiseux series in a
neighborhood of the point $z=\infty$. This balance and its solutions take the form
\begin{equation}
\begin{gathered}
 \label{Lienard_alt_dominant}
f_0s^{m+1}-(m+1)z=0,\quad s^{(k)}(z)=b_0^{(k)}z^{1/(m+1)},\quad k=1,\ldots, m+1,
\end{gathered}
\end{equation}
where $b_0^{(1)}$, $\ldots$, $b_0^{(m+1)}$ are distinct roots of the equation $f_0 b_0^{m+1}-(m+1)=0$.
The balance \eqref{Lienard_alt_dominant} is algebraic. We find $m+1$ distinct Puiseux series in a neighborhood of the point $z=\infty$.
The representation of $G(s,z)$ in $\mathbb{C}_{\infty}\{z\}$ is the following

\begin{equation}
\begin{gathered}
 \label{Lienard_alt_rep}
G(s,z)=\nu(z)\prod_{k=1}^{m+1}\{s-b_0^{(k)}z^{1/(m+1)}-\ldots\}^{n_k},
\end{gathered}
\end{equation}
where $\nu(z)\in\mathbb{C}[z]$ and $n_k=0$ or $n_k=1$. Since $G(s,z)$ should be a polynomial, we obtain
 from representation  \eqref{Lienard_alt_rep} that  the degree of $G(s,z)$ with respect to $s$ is either $0$
 ($n_k=0$, $k=1,\ldots,m+1$), or $m+1$ ($n_k=1$, $k=1,\ldots,m+1$).

\begin{table}[t]%[h]
        %\center
       \begin{tabular}[pos]{l l l}
        \hline
        \textit{Invariant algebraic curves} & \textit{Cofactors} & \textit{Parameters}\\
        \hline
        $ $ & $ $ & $ $\\
        $ y+\frac{\varepsilon}{3}x-\frac{\varepsilon\beta}{9}$ & $-3x^2-\beta x$ & $ \delta=\frac{\varepsilon\beta}{27}(\varepsilon-3\alpha)$,\\
            & $+\frac{\varepsilon}{3}-\alpha$ & $ \sigma=-\frac{\varepsilon}{18}(\beta^2+2\varepsilon-6\alpha)$\\
        $ y+x^3+(\alpha-\varepsilon) x+\frac{\delta}{\varepsilon}$ & $-\varepsilon$ & $ \beta=0 $, $\sigma=\varepsilon(\alpha-\varepsilon)$\\
        $ y^2+(x^2+\frac{4\varepsilon}{3})xy+\frac{\varepsilon}{3}x^4+\frac{4\varepsilon^2}{9}x^2$ & $-3x^2-\frac{8\varepsilon}{3}$ &
        $ \alpha=2\varepsilon$, $ \beta=0$, $\delta=0$, $\sigma=\frac{8\varepsilon^2}{9}$\\
        $ y^2+(x^2-\frac{4\varepsilon}{3})xy+\frac{\varepsilon}{3}x^4-\frac{8\varepsilon^2}{9}x^2+\frac{16}{27}\varepsilon^3$ & $-3x^2$ &
        $ \alpha=-\frac{2\varepsilon}{3}$, $ \beta=0$, $\delta=0$, $\sigma=-\frac{4\varepsilon^2}{3}$\\
        $ y^2+(x^3+\frac{28\varepsilon}{27}x+\frac{160i\varepsilon^{3/2}}{729})y+\frac{\varepsilon}{3}x^4$ & $-3x^2-\frac{64\varepsilon}{27}$ &
        $ \alpha=\frac{46\varepsilon}{27}$, $ \beta=0 $,   $\delta=\frac{32i\varepsilon^{5/2}}{243}$, \\
       $ \quad +\frac{8\varepsilon^2}{27}x^2+\frac{256i\varepsilon^{5/2}}{2187}x-\frac{80\varepsilon^3}{6561}$ & $ $ & $\sigma=\frac{52\varepsilon^2}{81}$\\
       $ y^2+(x^3+\frac{28\varepsilon}{27}x-\frac{160i\varepsilon^{3/2}}{729})y+\frac{\varepsilon}{3}x^4$ & $-3x^2-\frac{64\varepsilon}{27}$ &
        $ \alpha=\frac{46\varepsilon}{27}$, $ \beta=0 $, $\delta=-\frac{32i\varepsilon^{5/2}}{243}$, \\
       $ \quad +\frac{8\varepsilon^2}{27}x^2-\frac{256i\varepsilon^{5/2}}{2187}x-\frac{80\varepsilon^3}{6561}$ & $ $ & $\sigma=\frac{52\varepsilon^2}{81}$\\
       $ y^3+2(x^2+\frac{5\varepsilon}{3})xy^2+(x^4+\frac{11\varepsilon}{3}x^2$ & $-3x^2-\frac{25\varepsilon}{6}$ &
       $ \alpha=\frac{5\varepsilon}{2}$, $ \beta=0 $, $\delta=0$, $\sigma=\frac{25\varepsilon^2}{18}$\\
       $ \quad        +\frac{125\varepsilon^2}{36}) x^2y+\frac{\varepsilon}{3}x^7+\frac{11\varepsilon^2}{9}x^5+\frac{125\varepsilon^3}{108}x^3$ & $ $ & $ $\\
       $ y^3+(2x^2+\frac{15\varepsilon}{7})xy^2+(x^4+\frac{52\varepsilon}{21}x^2$ & $-3x^2-\frac{25\varepsilon}{7}$ &
       $ \alpha=\frac{40\varepsilon}{21}$, $ \beta=0 $, $\delta=0$, $\sigma=\frac{125\varepsilon^2}{147}$\\
        $ \quad        +\frac{75\varepsilon^2}{49}) x^2y+\frac{\varepsilon}{3}x^7+\frac{314\varepsilon^2}{551}x^5+\frac{125\varepsilon^3}{343}x^3$ & $ $ & $ $\\
       $ y^3+(2x^2-\frac{85\varepsilon}{27})xy^2+(x^6-\frac{76\varepsilon}{27}x^4$
       & $-3x^2-\frac{25\varepsilon}{27}$ &
       $ \alpha=-\frac{20\varepsilon}{27}$, $ \beta=0 $, $\delta=0$,\\
        $ \quad +\frac{1075\varepsilon^2}{729}x^2+\frac{400000\varepsilon^3}{531441})y  +\frac{\varepsilon}{3}x^7-\frac{14\varepsilon^2}{9}x^5$ & $ $ &
        $\sigma=-\frac{125\varepsilon^2}{81}$\\
         $      \quad    +\frac{15875\varepsilon^3}{6561}x^3-\frac{2000000\varepsilon^4}{1594323}x$ & $ $ & $ $\\
                              $ $ & $ $\\
                              \hline
    \end{tabular}
    \caption{Irreducible invariant algebraic curves of dynamical system \eqref{Lienard1_DS23} at  $e=0$, $\zeta=3$, and $\varepsilon\neq0$.} \label{T:FH_inv}
\end{table}

\textit{Proof of theorem \ref{T3_L23}.} We begin the prove by introducing the new variables $s$ and $z$ as described in the introduction.
The polynomial $h(x)$ takes the form $h(x)=x^3+\beta x^2/2+(\alpha-\varepsilon)x$. System \eqref{Lienard1_alt_DS} now is
\begin{equation}
\begin{gathered}
 \label{Lienard23_alt_sys}
s_t=z-s^3-\frac{\beta}{2} s^2+(\varepsilon-\alpha)s,\quad z_t=\frac{\beta\varepsilon}{2}s^2 +\{\varepsilon(\alpha-\varepsilon)-\sigma\}s-\delta-\varepsilon z.
\end{gathered}
\end{equation}
Invariant algebraic curves of system  \eqref{Lienard23_alt_sys} satisfy the equation
\begin{equation}
\begin{gathered}
 \label{Lienard23_alt_sys_EG}
\left(z-s^3-\frac{\beta}{2} s^2+(\varepsilon-\alpha)s\right)G_s+\left(\frac{\beta\varepsilon}{2}s^2 +\{\varepsilon(\alpha-\varepsilon)-\sigma\}s-\delta-
\varepsilon z\right)G_z=\lambda G,
\end{gathered}
\end{equation}
where it is straightforward to show that the highest--order coefficient of $G(s,z)$ with respect to $z$ is  $\mu(s)=\mu_0\in\mathbb{C}$
and the cofactor $\lambda$ takes the form $\lambda=A_2s^2+A_1s+A_0$, where $A_0$, $A_1$, $A_2\in\mathbb{C}$.
Without loss of generality, we shall set $\mu_0=1$. By direct substitution we verify that
there are no invariant algebraic curves that do not depend on $z$.

Further, constructing the Puiseux series, we find the representations of irreducible invariant algebraic curves
of system  \eqref{Lienard23_alt_sys} in the fields $\mathbb{C}_{\infty}\{z\}$ and $\mathbb{C}_{\infty}\{s\}$:
\begin{equation}
\begin{gathered}
 \label{Lienard23_alt_rep}
\mathbb{C}_{\infty}\{z\}:\, G(s,z)=\nu(z)\{s-b_0^{(1)}z^{1/3}-\ldots\}\{s-b_0^{(2)}z^{1/3}-\ldots\}\{s-b_0^{(3)}z^{1/3}-\ldots\};\hfill\\
\mathbb{C}_{\infty}\{s\}:\, G(s,z)=\left\{z-s^3-\frac{\beta}{2}s^2-\left(\alpha-\frac{4}{3}\varepsilon\right)s-\frac{\beta\varepsilon}9-\ldots\right\}^k
\prod_{j=1}^{N-k}\{z-a_0^{(j)}-\ldots\}.\hfill
\end{gathered}
\end{equation}
Here $N\in\mathbb{N}$, $\nu(z)\in\mathbb{C}[z]$, $k=0$ or $k=1$.  The Puiseux series 
\begin{equation}
\begin{gathered}
 \label{Lienard23_alt_rep_PS1}
(I):\quad z(s)=s^3+\frac{\beta}{2}s^2+\left(\alpha-\frac{4}{3}\varepsilon\right)s+\frac{\beta\varepsilon}9+\sum_{l=1}^{\infty}d_ls^{-l}.
\end{gathered}
\end{equation}
possesses uniquely determined coefficients. The Puiseux series 
\begin{equation}
\begin{gathered}
 \label{Lienard23_alt_rep_PS2}
(II):\quad z(s)=a_0+\frac{\varepsilon(\alpha-\varepsilon)-\sigma}{s}-\frac{\varepsilon a_0+\delta}{2s^2}+\sum_{l=3}^{\infty}a_ls^{-l}.
\end{gathered}
\end{equation}
has an arbitrary coefficient $a_0$ and exists whenever $\beta=0$. Thus if $\beta\neq0$ we obtain only one irreducible invariant algebraic curve
$G(s,z)=z-s^3-(\beta s^2)/2-\{\alpha-(4\varepsilon/3)\}s-(\beta\varepsilon)/9$ existing whenever the corresponding series terminates at zero term.
This gives the following restrictions on the parameters $ \delta=\varepsilon\beta(\varepsilon-3\alpha)/27$, 
$ \sigma=-\varepsilon(\beta^2+2\varepsilon-6\alpha)/18$.
Further, we set $\beta=0$ and calculate several first coefficients of the
Puiseux series under consideration. Requiring that the non--polynomial part of the second
 expression in \eqref{Lienard23_alt_rep}  be zero, we obtain necessary conditions for invariant algebraic curves to exist. 
Since $G(s,z)$ is irreducible,  we see that $N=1$ whenever $k=0$. 
 Setting $N=1$ and $k=0$, 
 we find  conditions for the series of type $(II)$ to terminate at zero term. As a result we obtain $\sigma=\varepsilon(\alpha-\varepsilon)$
 and $G(s,z)=z+\delta/\varepsilon$. Considering the case $k=1$, we calculate the coefficients at $z^{N-1}s^{-l}$, $l\in\mathbb{N}$ in the representation 
 of $G(s,z)$ in the field $\mathbb{C}_{\infty}\{s\}$. This yields the following 
 relations
  \begin{equation}
\begin{gathered}
 \label{Lienard23_alt_rep_rel1}
d_l+\sum_{j=1}^{N-1} a^{(j)}_l=0,\quad l\in\mathbb{N}.
\end{gathered}
\end{equation}
 It is convenient to introduce the variables $C_m=\{a^{(1)}\}^m+\ldots+\{a^{(N-1)}\}^m$, where $m\in\mathbb{N}$. In practice we need  thirteen
 relations given by  \eqref{Lienard23_alt_rep_rel1} with $1\leq l\leq13$ in order to find all other irreducible invariant algebraic curves.
 Sufficiency we verify by direct substitution into
 equation \eqref{Lienard23_alt_sys_EG}. Finally, we return to the original variables $x$ and $y$. The results are gathered in table \ref{T:FH_inv}.

\textit{Remark.} Since the degree with respect to $s$ of irreducible invariant algebraic curves of system  \eqref{Lienard23_alt_sys}
is either $0$ or $3$ (theorem \ref{T2_L_alt}),
one can use the method of undetermined coefficients to obtain the curves explicitly, see \cite{Demina05}.

\section{Conclusion}

In this article  a new method of finding
invariant algebraic curves of a polynomial dynamical system has been introduced. The structure of invariant algebraic curves
of the Li\'{e}nard differential system  \eqref{Lienard_gen} with $\deg g=\deg f+1$ has been established. Using the new method all irreducible invariant algebraic curves
of the Li\'{e}nard differential system  \eqref{Lienard_gen} with $\deg f=2$ and $\deg g=3$ have been obtained.

\section{Acknowledgments}

The author is grateful to the reviewer for helpful comments and constructive suggestions.

\bibliography{ref}

\end{document}